\documentclass[11pt,onecolumn,a4paper]{article}

\usepackage[latin1]{inputenc}
\usepackage[T1]{fontenc}
\usepackage{cite}
\usepackage[dvips]{epsfig}
\usepackage[dvips]{color}
\usepackage{amsmath}
\usepackage{graphicx}
\usepackage{bm}
\usepackage{mathrsfs}
\usepackage{float}
\usepackage{slashed}
\usepackage[english]{babel}
\usepackage{subfigure}

\begin{document}

\title{\textbf{\Large Circularly polarized waves in a plasma with vacuum
polarization effects}}
\author{J. Lundin, L. Stenflo, G. Brodin, M. Marklund and P. K. Shukla \\
\textit{Department of Physics, Ume\aa\ University, SE-901 87 Ume\aa, Sweden}}
\maketitle

\begin{abstract}
The theory for large amplitude circularly polarized waves propagating along
an external magnetic field is extended in order to include also vacuum
polarization effects. A general dispersion relation, which unites previous results, is derived.
\end{abstract}

Circularly polarized waves can be \emph{exact} solutions of the fluid and Maxwell
equations (e.g.\ Refs.\ \cite{Stenflo-1974, Stenflo-1976}). Thus, it is
possible to study \emph{large} amplitude waves propagating along an external
magnetic field in a plasma, taking also relativistic effects into account.
Here we shall demonstrate that it is comparatively easy to include
quantum electrodynamical field (QED) effects \cite{Brodin-2001,Marklund-2006} in such an analysis, also for
the general case with a multicomponent plasma.

Let us start with the equations of continuity and momentum for each species
(denoted by index $s$) in a cold, multi-component, relativistic, magnetized
plasma 
\begin{equation}
\partial _{t}N_{s}+\nabla \cdot \left( N_{s}\mathbf{v}_{s}\right) =0,
\end{equation}
and 
\begin{equation}\label{eq:eq2}
\partial _{t}\mathbf{p}_{s}+\mathbf{v}_{s}\cdot \nabla \mathbf{p}%
_{s}=q_{s}\left( \mathbf{E}+\mathbf{v}_{s}\times \mathbf{B}\right) .
\end{equation}
Here $N$ is the particle density, $\mathbf{v}$ the velocity, $q$ the charge, 
$m$ the rest mass, $\mathbf{p}=m\mathbf{v}/(1-v^{2}/c^{2})^{1/2}$ the momentum and $c$ the speed of light in vacuum. Furthermore, we
use the Maxwell's equations for the electric and magnetic fields $\mathbf{E}$
and $\mathbf{B}$, i.e.
\begin{equation}
\nabla \times \mathbf{E}=-\partial _{t}\mathbf{B},  \label{eq:M3}
\end{equation}
and 
\begin{equation}
\nabla \times \mathbf{B}-c^{-2}\partial _{t}\mathbf{E}=\mu
_{0}\sum_{s}N_{s}q_{s}\mathbf{v}_{s}+\mu _{0}\mathbf{J}_{\text{vac}},
\label{eq:M4}
\end{equation}
where the vacuum current $\mathbf{J}_{\text{vac}}$ (after having corrected an obvious misprint in the
definition of the coupling constant in Ref. \cite{Brodin-2001}) is
given by (e.g.\ Refs.\ \cite{DiPiazza-2007,Soljacic-2000})
\begin{eqnarray}
\mathbf{J}_{\text{vac}} &=&-2\kappa \varepsilon _{0}^{2}\left\{ c^{2}\nabla
\times \left[ 2(E^{2}-c^{2}B^{2})\mathbf{B}-7(\mathbf{E}\cdot \mathbf{B})%
\mathbf{E}\right] \right.  \notag   \\
&&-\left. \partial _{t}\left[ 2(E^{2}-c^{2}B^{2})\mathbf{E}%
+7c^{2}(\mathbf{E}\cdot \mathbf{B})\mathbf{B}\right] \right\} \label{eq:Jvac},
\end{eqnarray}
where $\kappa =2\alpha ^{2}\hbar ^{3}/45m_{e}^{4}c^{5}$ is the nonlinear
coupling constant and $\alpha =e^{2}/4\pi \hbar c\varepsilon _{0}$ is the
fine structure constant. Here $e$ $(=-q_e)$ is the elementary charge, $m_{e}$ the
electron rest mass, $\varepsilon _{0}$ the vacuum permittivity, $\mu _{0}$
the vacuum permeability and $\hbar $ is $2\pi $ times the Planck constant.
The expression (\ref{eq:Jvac}) for the vacuum current is valid for field
strengths lower than the critical field strength, $E_{\text{crit}%
}=m_{e}^{2}c^{3}/\hbar e$, and for frequencies lower that the Compton
frequency, $\omega _{e}=m_{e}c^{2}/\hbar $. These equations have been recently used
by Di Piazza et al.\ \cite{DiPiazza-2007}. 

We now consider the
propagation of a circularly polarized wave along an external constant
magnetic field $B_{0z}\hat{\mathbf{z}}$. Following Ref.\ \cite{Stenflo-1976}
we first note that $E^{2}$ as well as $B^{2}$ are then constants and that $%
\mathbf{E}\cdot \mathbf{B}=0$. This means that 
\begin{equation}\label{eq:Jvacuum}
\mathbf{J}_{\text{vac}}=-4\kappa \varepsilon _{0}^{2}\left(
E^{2}-c^{2}B^{2}\right) \left( c^{2}\nabla \times \mathbf{B}-%
\partial _{t}\mathbf{E}\right).
\end{equation}
Inserting Eq.\ (\ref{eq:Jvacuum}) into Eq.\ (\ref{eq:M4}) we thus find our new
equation 
\begin{subequations}\label{eq:MQED4}
\begin{eqnarray}
	\nabla \times \mathbf{B}-c^{-2}\partial _{t}\mathbf{E}=\mu _{0}%
	\mathbf{J}_{\text{eff}},
\end{eqnarray}
where 
\begin{eqnarray}
	\mathbf{J}_{\text{eff}}=\sum_{s}N_{s}q_{s}\mathbf{v}_{s}\left\{ 1-%
	\frac{2\alpha }{45\pi }\left[ (n^{2}-1)\left( \frac{E_{0\bot }}{E_{\text{crit%
	}}}\right) ^{2}+\left( \frac{cB_{0z}}{E_{\text{crit}}}\right) ^{2}\right]
	\right\}^{-1} .
\end{eqnarray}
\end{subequations}
Here $n=kc/\omega $ is the index of refraction, and $E_{0\bot }$ is the
electric field amplitude of our right hand circularly polarized wave $\mathbf{E}%
=E_{0\bot }\left[ \cos \left( \omega t-kz\right) \hat{\mathbf{x}}+\sin
\left( \omega t-kz\right) \hat{\mathbf{y}}\right] $. The rest of the
calculations are now straightforward. Introducing the symbols $E_{\pm
}=E_{x}\pm iE_{y}$, $B_{\pm }=B_{x}\pm iB_{y}$ and $v_{\pm }=v_{x}\pm iv_{y}$
we thus adopt the analysis of Ref.\ \cite{Stenflo-1976}, noting that 
\begin{subequations}\label{eq:Stenflo}
\begin{eqnarray}
	E_{\pm }=E_{0\bot }\exp \left( \pm i\omega t\mp ikz\right) ,
\end{eqnarray}
\begin{eqnarray}
	B_{\pm }=\pm i\frac{k}{\omega }E_{\pm },
\end{eqnarray}
and 
\begin{eqnarray}
	v_{\pm s}=\mp \frac{iq_{s} }{\gamma
	_{s}m_{s} \left( \omega +\omega _{cs}\right) }E_{\pm },
\end{eqnarray}
\end{subequations}
where  $\omega _{cs}=q_{s}B_{0z}/\gamma _{s}m_{s}$ is the
relativistic gyrofrequency of particle species $s$ and $\gamma=(1-v_{0}^{2}/c^{2})^{-1/2}$ is
the Lorentz factor which in this case is a constant with $%
v_{0}^{2}=v_{+}v_{-}$. Usually, we define the relativistic plasma
frequency as $\omega _{ps}=\left( q_{s}^{2}N_{0s}/\gamma _{s}m_{s}\varepsilon
_{0}\right) ^{1/2}$. Here, however, instead of $\omega _{ps}$ we will in the
dispersion relation find the effective plasma frequency 
\begin{equation*}
\Omega _{ps}=\omega _{ps}\left\{ 1-\frac{2\alpha }{45\pi }\left[
(n^{2}-1)\left( \frac{E_{0\bot }}{E_{\text{crit}}}\right) ^{2}+\left( \frac{%
cB_{0z}}{E_{\text{crit}}}\right) ^{2}\right] \right\}^{-1/2}.
\end{equation*}
By combining Eqs.\ (\ref{eq:MQED4}) and (\ref{eq:Stenflo}) we obtain the
dispersion relation 
\begin{equation}\label{eq:disprel}
n^{2}-1=-\sum_{s}\frac{\omega _{ps}^{2}}{\omega\left( \omega +\omega _{cs}\right) }+\frac{2\alpha 
}{45\pi }\left[ \left( n^{2}-1\right) \left( \frac{E_{0\bot }}{E_{\text{crit}%
}}\right) ^{2}+\left( \frac{cB_{0z}}{E_{\text{crit}}}\right) ^{2}\right]
\left( n^{2}-1\right) .
\end{equation}
The theory for a left hand circularly polarized wave is quite similar and leads to the same dispersion relation but with $\omega_{cs} \rightarrow -\omega_{cs}$.

We note that Eq.\ (\ref{eq:disprel}) agrees with Eq.\ (13) in Ref.\ \cite
{Marklund-2005} in the special case of an ultra-relativistic
electron-positron plasma, and with Eq.\ (9) in Ref.\ \cite{Marklund-2005b}
for a dusty plasma (if we correct for a misprint in Refs. \cite
{Marklund-2005} and \cite{Marklund-2005b}; $4\alpha $ should be $2\alpha $).
Eq.\ (\ref{eq:disprel}) is exact within the basic model we have used in the present
paper.

As $\alpha E^2_{0\bot}/E_{\text{crit}}^2$ in general is a very small
parameter, we can next expand the solution of Eq.\ (\ref{eq:disprel}) to
obtain 
\begin{eqnarray}  \label{eq:disprelexpansion}
n^2=1-\sum_s \frac{\omega_{ps}^2}{\omega\left(\omega+\omega_{cs}\right)}-\frac{2\alpha}{45\pi}\left(\frac{%
cB_{0z}}{E_{\text{crit}}}\right)^2\sum_s\frac{\omega_{ps}^2}{\omega\left(\omega+\omega_{cs}\right)} 
\notag \\
+\frac{2\alpha}{45\pi}\left(\frac{E_{0\bot}}{E_{\text{crit}}}%
\right)^2\left[\sum_s\frac{\omega_{ps}^2}{\omega\left(\omega+\omega_{cs}\right)}\right]^2,
\end{eqnarray}
which agrees with the result of Ref.\ \cite{DiPiazza-2007} in the limit $%
B_{0z}=0$.

New modes in unmagnetized QED plasmas have been reported previously \cite
{Stenflo-2005}. The calculation were then extended to magnetized QED plasmas
with ultrarelativistic ions. Such plasmas can support amplitude-dependent
electromagnetic ion waves \cite{Stenflo-1979} 
\begin{equation*}
\omega \approx \frac{k^{2}c^{2}}{\omega _{pi}^{2}}\omega _{Ei},
\end{equation*}
where $\omega _{pi}$ is the ion plasma frequency and $\omega _{Ei}=eE_{0\bot
}/cm_{i}$, as well as new modes in a dust-ion plasma \cite{Stenflo-2001},
and QED effects can then play a key role \cite{Marklund-2005b,
Stenflo-2001}. Di Piazza et al.\ demonstrated in Ref.\ \cite{DiPiazza-2007}
that the QED effects can be important in wave propagation studies, even for
the case where $B_{0z}=0$. They showed that the last term in Eq.\ (\ref
{eq:disprelexpansion}) can be larger than the sum of the first two terms on
the right hand side of Eq.\ (\ref{eq:disprelexpansion}) near the plasma
cut-off frequency. Thus, they demonstrated enhancement of the vacuum polarization
effects when the frequency is close to $\omega _{p}$. Our dispersion
relation shows that there is no analogous enhancement of the QED
contribution from the external magnetic field near the plasma cut-off
frequency. This is because the 2nd and the 3rd term on the right hand
side of Eq.\ (\ref{eq:disprelexpansion}) have the same sign. Thus, the QED
contribution from the external magnetic field will always remain smaller
than the classical contribution in Eq.\ (\ref{eq:disprelexpansion}) unless $%
cB_{0z}\gg E_{\text{crit}} $, but this condition lies outside the
validity of our model. The QED contribution from the external magnetic field
may still be larger than that of the last term in Eq.\ (\ref
{eq:disprelexpansion}), however, in particular if $cB_{0z}>E_{0\bot }$, $\omega
_{p}<\omega $ and if $\left( \omega+\omega _{cs}\right) $ is not close to resonance. But in this regime there is no enhancement of the
vacuum polarization effects as described in Ref.\ \cite{DiPiazza-2007}.

It is also interesting to note that the effective plasma frequency $\Omega
_{p}$ is different from $\omega _{p}$. For example if $k\approx 0$ we have 
\begin{equation*}
\frac{\Omega _{ps}}{\omega _{ps}}=\left[ 1+\frac{2\alpha }{45\pi E_{\text{%
crit}}^{2}}\left( E_{0\bot }^{2}-c^{2}B_{0z}^{2}\right) \right] ^{-1/2}.
\end{equation*}
However, for field strengths available in laboratory experiments, this
ratio is close to unity. 

Another example of interest concerns the free electron laser (FEL) \cite
{Kwan-1977,Goldring-1985,Freund1986}. Considering magnetostatic wigglers, i.e.\ letting $\omega
\rightarrow 0$, in a one-component electron plasma with drift velocity $v_{ze}$ and gyrofrequency $\omega_{ce}$, the
dispersion relation (\ref{eq:disprel}), using the laboratory frame rather than the electron rest frame, becomes
\begin{equation}\label{eq:FEL-DR}
k^{2}c^{2}\left[ 1-\frac{2\alpha }{45\pi }\left( \frac{cB_{\mathrm{tot}}}{E_{%
\text{crit}}}\right) ^{2}\right] =\frac{\omega _{pe}^{2}kv_{ze}}{
\omega _{ce}-kv_{ze} }
\end{equation}
where we have introduced the total magnetic field $B_{\mathrm{tot}}=\sqrt{%
B_{0\bot }^{2}+B_{0z}^{2}}$ and the magnetic field amplitude $B_{0\bot }$ of
the wiggler field. For $cB_{\mathrm{tot}}\ll E_{%
\text{crit}}$, Eq. (\ref{eq:FEL-DR}) agrees for example with (8) in Ref.\ \cite{Freund1986}. Thus our formula (\ref{eq:FEL-DR})
generalizes the FEL dispersion relation to account for field strengths
approaching the Schwinger critical field. While current laboratory
applications have a long way to go in order to have fields in
this range, there nevertheless exist schemes attempting to reach such
regimes \cite{Ringwald2001}. We note that the particle orbits in FELs are nontrivial \cite{Goldring-1985, Freund1986} and that the helical equilibria may not be stable \cite{Goldring-1985}. In the example above we have also supposed the presence of a stationary background assuring neutrality, which means that there are several limiting assumptions concerning forthcoming experiments \cite{DiPiazza-2007}. Thus we hope that the readers will find interest in further extending the basic theory of Ref.\ \cite{Stenflo-1976}.

In the present manuscript we have derived the dispersion relation for large
amplitude circularly polarized electromagnetic waves propagating parallel
to an external magnetic field in a multicomponent plasma. Our dispersion
relation (\ref{eq:disprel}) accounts for vacuum polarization and
magnetization (both due to the external field and the wave field), fully
relativistic quiver velocities and general drift velocities of the plasma
species, and applies for arbitrary ratios of the characteristic frequencies
of the problem. As a consequence, such different waves as Alfvén modes,
whistler modes and large amplitude laser modes are covered by the general
theory. Considering specific regimes, Eq. (\ref{eq:disprel}) reduces to
several special cases found in the literature. Besides uniting the results
of many previous works in a single formalism, the results presented here can
find applications to wave propagation in astrophysical contexts, e.g. in
magnetar atmospheres \cite{Baring1998}, as well as in the next generation of
FELs \cite{Ringwald2001}.

We consider the present work as a necessary prerequisite to a more complete
theory which also includes a full stability analysis \cite{Stenflo-1976}.
Thus it has for example been shown \cite{Stenflo-1974} that a single cold
electron beam in an unmagnetized plasma is unstable if $k=0$ and if
\begin{equation}
\left( 1+\frac{\omega _{pe}^{2}}{K^{2}c^{2}}\right) ^{-1}<\frac{v_{0}^{2}}{%
c^{2}+v_{0}^{2}}<\frac{K^{2}c^{2}}{\Omega ^{2}}
\end{equation}
where $\Omega $ and $K$ are the frequency and wavenumber of the
perturbations. The vacuum current will however modify all stability criteria.

\end{document}